\documentclass[pre, reprint, showpacs, superscriptaddress]{revtex4-1}
\usepackage{amsmath, graphicx,amssymb,color, bm}
\begin{document}
\title{Kinetics of symmetry and asymmetry in a phase-separating bilayer membrane}

\date{\today}

\author{J.~J.~Williamson}
\email{johnjosephwilliamson@gmail.com}
\author{P.~D.~Olmsted}
\email{pdo7@georgetown.edu}
\affiliation{Department of Physics, Institute for Soft Matter Synthesis and Metrology, Georgetown University, 37th and O Streets, N.W., Washington, D.C. 20057, USA}

\begin{abstract} 
We simulate a phase-separating bilayer in which the leaflets experience a direct coupling favouring local compositional symmetry (``registered'' bilayer phases), and an indirect coupling due to hydrophobic mismatch that favours strong local asymmetry (``antiregistered'' bilayer phases). For wide ranges of overall leaflet compositions, multiple competing states are possible. For estimated physical parameters, a quenched bilayer may first evolve toward a metastable state more asymmetric than if the leaflets were uncorrelated;\ subsequently, it must nucleate to reach its equilibrium, more symmetric, state. These phase-transition kinetics exhibit characteristic signatures through which fundamental and opposing inter-leaflet interactions may be probed. We emphasise how bilayer phase diagrams with a separate axis for each leaflet can account for overall and local symmetry/asymmetry, and capture a range of observations in the experiment and simulation literature. 
\end{abstract}
\color{black}

\maketitle

\section{\label{sec:intro}Introduction}

Phase separation within mixed bilayers is intimately linked to the symmetry or otherwise of their separate, yet coupled, leaflets \cite{Williamson2014, Collins2008, Korlach1999, Dietrich2001, May2009,Putzel2011,Funkhouser2013, Allender2006, Putzel2008, Wagner2007, Hirose2009}. The \textit{overall} compositions of the leaflets may by asymmetric;\ for instance, one leaflet containing an equal mixture of saturated and unsaturated lipids and the other predominantly unsaturated lipids. The phase behaviour of such bilayers differs dramatically from those whose overall leaflet compositions are symmetric \cite{Collins2008}. Alternatively, even if the overall compositions are symmetric, as is common in model systems, phase separation can lead to either registered or antiregistered phases, and thus \textit{local} symmetry or asymmetry \cite{Perlmutter2011, Reigada2015}.

There is evidence that bilayers experience competing inter-leaflet coupling effects. Many observations have shown separation into registered (R) bilayer phases, those which are locally symmetric, comprising leaflets with the same phase and composition \cite{Korlach1999, Dietrich2001, Collins2008, Reigada2015}.
This implies direct inter-leaflet composition coupling via a mismatch energy per area $\gamma$ \cite{May2009, Putzel2008, Putzel2011, Risselada2008, Polley2013}. Such a direct coupling is also used to explain apparent induction or suppression of domains in one leaflet by the other, in cases where overall leaflet compositions are asymmetric \cite{Collins2008, Lin2015}.
Estimates from theory \cite{Putzel2011, May2009}, simulation \cite{Pantano2011, Risselada2008, Polley2013} and, more recently, experiment \cite{Blosser2015} vary over approximately $\gamma = 0.01 - 1\, k_\textrm{B}T\textrm{nm}^{-2}$. 
Conversely, lipid hydrophobic length mismatch indirectly couples the leaflets via their combined hydrophobic thickness. This effect promotes strong local asymmetry, i.e., antiregistered (AR) bilayer phases, to give more uniform bilayer thickness and avoid the elastic cost \cite{Galimzyanov2015} that would otherwise result from hydrophobic mismatch \cite{Perlmutter2011, Reigada2015, Stevens2005, Zhang2004, Zhang2007, Bossa2015}. 

A microscopic model \cite{Williamson2014} incorporating these effects predicts a kinetic competition of R and AR bilayer phases. Linear stability analysis in physical parameter ranges predicts that instabilities to forming registered or antiregistered phases can be of comparable strength, so that which occurs first after a quench is determined by molecular properties such as tail length or unsaturation \cite{Williamson2014}. If a metastable state forms first, nucleation is required to subsequently initiate the equilibrium (typically registered) phases \cite{Williamson2015a}.

\begin{figure}[floatfix]
\includegraphics[width=8.0cm]{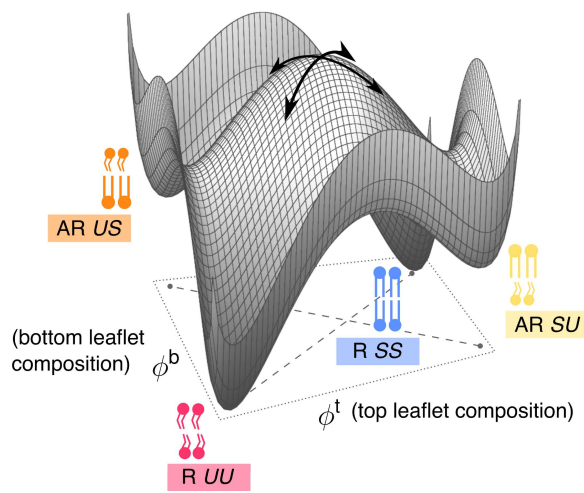}
\caption{\label{landscape}(Color online) Schematic free-energy landscape $f(\phi^\textrm{t},\phi^\textrm{b})$, which determines available phase coexistences \cite{Williamson2014}. Cartoons show the dominant transbilayer arrangement of model $S$ and $U$ species in R and AR bilayer phases. Curved arrows represent competing linear instabilities for a bilayer of $(0.5,0.5)$ overall composition. 
}
\end{figure}

In this paper we examine the resulting kinetics by direct simulation of the microscopic lattice model \cite{Williamson2014}. Multiple phase coexistences involving registered and antiregistered phases compete over wide regions of the phase diagram. Even in cases where the given overall leaflet compositions prevent the attainment of either ``perfect'' local symmetry or asymmetry throughout the bilayer, the bilayer may still select between states with greater or lesser amounts of registered and antiregistered phases than the uncorrelated case of fully-independent leaflets.

We find kinetic signatures of an Ostwald stage rule by which a bilayer progresses through metastable states on its way to equilibrium. For example, the nonmonotonic evolution of a parameter describing the degree of molecular transbilayer symmetry could be measured in molecular simulation or in experiment. We emphasise the utility of bilayer phase diagrams with a separate composition axis for each leaflet \cite{Collins2008, Putzel2008, May2009}. We show how these relate to available experimental results, and to conventional phase diagrams that do not account for transbilayer symmetry/asymmetry. 

\section{Lattice model} \label{sec:lattice}

The model introduced in \cite{Williamson2014} describes a local bilayer patch as $N$ sites on a square lattice of spacing $a \sim 0.8\,\textrm{nm}$, where each site contains top and bottom leaflet lipids. Each lipid has a hydrophobic length $\ell^\textrm{t(b)}_i$, from which we define the bilayer thickness 
\begin{equation}
d_i \equiv \ell^\textrm{t}_i + \ell^\textrm{b}_i~,
\end{equation}
\noindent and leaflet thickness difference 
\begin{equation}
\Delta_i \equiv \ell^\textrm{t}_i - \ell^\textrm{b}_i~.
\end{equation}
\noindent Model species $S$ and $U$ represent either saturated and unsaturated lipids, or the more ordered (liquid-ordered $L_{o}$ or gel) versus liquid-disordered ($L_{d}$) states of a ternary mixture (Section~\ref{sec:qual}).

Defining $\hat{\phi}^\textrm{t(b)}_i = 1$ if the top (bottom) of site $i$ contains an $S$ lipid, $\hat{\phi}^\textrm{t(b)}_i = 0$ if $U$, the Hamiltonian is
\begin{align}\label{eqn:fun4}
H = &\sum_{<i,j>} ( V_{\hat{\phi}_i^\textrm{t} \hat{\phi}_j^\textrm{t}}  +  V_{\hat{\phi}_i^\textrm{b} \hat{\phi}_j^\textrm{b}}) 
+ \sum_{<i,j>} \tfrac{1}{2}\tilde{J} (d_i - d_j)^2 \notag \\
&+  \sum_{i} \tfrac{1}{2}B (\Delta_i)^2 
+  \sum_{i} \tfrac{1}{2}\kappa \left(( \ell^\textrm{t}_i - \ell_0^{\textrm{t}i})^2 + ( \ell^\textrm{b}_i - 
\ell_0^{\textrm{b}i})^2                      \right)~,
\end{align}
\noindent where species-dependent ideal (i.e., preferred) hydrophobic lengths are $\ell_0^{\textrm{t(b)}i} = \ell_{S0}$ for an $S$ lipid at the top (bottom) of site $i$, or $\ell_{U0}$ for $U$ \footnote{With varying temperature, a given lipid may convert discontinuously between more and less-ordered states. This could be handled by using higher-order terms within the $\kappa$ term of Eq.~\ref{eqn:fun4} to yield multiple energy minima as a function of $\ell^\textrm{t(b)}_i$ for a given species \cite{Komura2004}. However, we consider a mixed system at a fixed temperature far from the melting temperature $T_\textrm{m}$ of the high-$T_\textrm{m}$ and low-$T_\textrm{m}$ pure lipids, such that the assumption of a conserved population of $S$ and $U$ lipids, each having a single preferred degree of tail length/ordering, is reasonable.}. 
Each site is pairwise registered ($SS$ or $UU$) or antiregistered ($SU$ or $US$). 
An R bilayer phase is one dominated by $SS$ or $UU$ sites such that, microscopically, most lipids face one of the same type in the apposed leaflet, while in an AR phase most lipids face one of the opposite type. 

The Ising or Flory-like $V_{uv}$ captures length-independent interactions occurring among neighbours within each leaflet, arising from, e.g., headgroup interactions. Its strength is quantified by $V \equiv V_{10} - \tfrac{1}{2}(V_{00} + V_{11})$. In mean-field, phase separation of the Ising model (i.e., if no other interactions were present) requires $V > 0.5\,k_\textrm{B}T$, while $V > 0.88\,k_\textrm{B}T$ is required in the presence of fluctuations, as in a simulation \cite{Huang1987}.

The ``direct'' inter-leaflet coupling $B$ is suggested by observations of registered domains in experiment and simulation \cite{Korlach1999, Dietrich2001, Collins2008, Perlmutter2011, Reigada2015}, as well as the apparent ``induction'' of domains in one leaflet by those in the other \cite{Lin2015} (see Appendix~\ref{app:details}). These observations imply a direct composition coupling between the leaflets, resulting in a mismatch energy per area for local compositional asymmetry \cite{Pantano2011, Risselada2008, Polley2013, Putzel2011, May2009, Blosser2015}. The direct coupling is expected to depend on tail structural features \cite{Putzel2011, May2009};\ here, tail structure is implicitly mapped to tail length \cite{Komura2004}, and the $B$ term in our model promotes pairwise R by penalising tail length mismatch across the midplane. 
The exact mechanisms (e.g., transmidplane tail interactions, leaflet curvature) are not crucial, though, and for comparison with the literature we can estimate an effective conventional inter-leaflet mismatch energy $\gamma$ (shown later on an axis of Fig.~\ref{stabdiag}) \cite{Pantano2011, Risselada2008, Polley2013, Putzel2011, May2009, Blosser2015}. 

The hydrophobic ``indirect'' coupling $\tilde{J}$ promotes pairwise AR, by penalising mismatches in the bilayer thickness profile \cite{Stevens2005, Perlmutter2011, Reigada2015}. We define $J \equiv 4\tilde{J}$, which appears in the mean-field approximation of Eq.~\ref{eqn:fun4} from which the coarse-grained free-energy density (Fig.~\ref{landscape}) is derived \cite{Williamson2014}. $\kappa$ can be related to the area compression modulus $\kappa_A$ by mapping length changes to area changes assuming constant volume of the hydrophobic tails (described in \cite{Williamson2014}). 
It quantifies the penalty for deviation from $ \ell_0^{\textrm{t(b)}i}$, so that smaller values soften both inter-leaflet couplings. While the interplay of $J$ and $\kappa$ gives a reasonable energy scale for the line tension associated with hydrophobic mismatch \cite{Williamson2015a}, it does not capture the detail of the bilayer's internal elastic deformations \cite{Galimzyanov2015}. We will explore implications of the findings of \cite{Galimzyanov2015} in an upcoming Comment on that work \cite{Comment}.

For simplicity the parameters $B$, $J$, and $\kappa$ are assumed to be the same for both lipid species. This will not affect the key qualitative features that are of interest here, but relaxing this assumption could lead to interesting behaviour in specific systems.

$\Delta_0 \equiv \ell_{S0} - \ell_{U0}$ quantifies the intrinsic mismatch in both preferred tail length and structure between species;\ although cast as a tail length difference, it works with \textit{both} $J$ and $B$ to control the strength of the indirect and direct inter-leaflet couplings. We model varying the tail hydrophobic length mismatch \textit{alone} by varying $J$, while varying $B$ alone corresponds to changing the degree of tail structure mismatch (for instance by increasing the difference in unsaturation). A reference total thickness $d_0 \equiv \ell_{S0} + \ell_{U0}$ may be defined, but in the absence of an external field acting on bilayer thickness the absolute values of $\ell_{S0}$ and $\ell_{U0}$ are irrelevant and only their difference plays a role. 

Recent work employed a model capturing similar competing inter-leaflet coupling effects to those studied here \cite{Bossa2015}. The authors successfully analysed local correlations and clustering via a quasi-chemical approximation, in the context of a laterally homogeneous bilayer rather than one undergoing phase separation.

\subsection{Mean-field free energy landscape}

Eq.~\ref{eqn:fun4} leads to the mean-field local free-energy density $f(\phi^\textrm{t},\,\phi^\textrm{b},\,\bar{d},\,\overline{\Delta})$ \cite{Williamson2014} as a function of average leaflet compositions in the local patch  
\begin{equation}\label{eqn:localphi}
\phi^\textrm{t(b)} \equiv \frac{N_S^\textrm{t(b)} }{N}~,
\end{equation}
where there are $N_S^\textrm{t(b)}$ top (bottom)-leaflet $S$ lipids in the local patch,
and of thickness variables
\begin{align}
\bar{d} &\equiv \frac{1 }{N}\sum d_i ~, \notag \\
\overline{\Delta} &\equiv \frac{1  }{N}\sum \Delta_i~,
\end{align}
\noindent where the sums are over the local patch. The local free-energy density as a function of local composition is
\begin{align}
f^\textrm{[ann.]}(\phi^\textrm{t},\phi^\textrm{b}) \equiv  f (\phi^\textrm{t},\,\phi^\textrm{b},\,\bar{d}^\textrm{[ann.]},\,\overline{\Delta}^\textrm{[ann.]})~.
\end{align}
\noindent The label $\textrm{[ann.]}$ (annealed) indicates equilibration of the local thickness variables at given local compositions \cite{Williamson2014} and will be omitted hereafter. Each leaflet is described by a separate composition variable, and bilayer phase behaviour can then be calculated in the space of $(\phi^\textrm{t},\phi^\textrm{b})$ (Section~\ref{sec:bilphasdiag}), similarly to \cite{Putzel2008, May2009}. An illustrative free-energy landscape is shown in Fig.~\ref{landscape}. 

The model species have the same molecular area \cite{Williamson2014}. Breaking this assumption, or otherwise breaking the symmetry of the free-energy landscape (e.g., allowing parameters such as $\kappa$ to vary between species) should not qualitatively affect the results. (However, in experiment or molecular simulation one would typically interpret $\phi^\textrm{t(b)}$ as \textit{area} fractions, which are equivalent to \textit{number} fractions in our model. For instance, an ``equimolar mixture'' corresponds to an ``equal area fractions mixture'' in a real system.) Such a change would remove the symmetry of the free energy (Fig.~\ref{landscape}) and phase diagram (Fig.~\ref{phasediag}b,c) under reflection through the $\phi^\textrm{b}=1-\phi^\textrm{t}$ line. A separate consideration is that biological membranes with, e.g., different solvent environments for each leaflet could break the top-bottom symmetry, thus removing the symmetry under reflection through the $\phi^\textrm{b}=\phi^\textrm{t}$ line.

\subsection{Bilayer phase diagrams and coexistence}\label{sec:bilphasdiag}

For the simulations presented below, the bilayer size is $\mathcal{L}^2 = 200^2 = \mathcal{N}$. Script letters refer to the entire bilayer, while $N$ used above described a local, homogeneous bilayer patch. Similarly, we explicitly define the \textit{overall} leaflet compositions 
\begin{equation}\label{eqn:overallphi}
\Phi^\textrm{t(b)} \equiv \frac{\mathcal{N}_S^\textrm{t(b)} }{\mathcal{N}}~,
\end{equation}
\noindent as distinct from the local compositions defined in Eq.~\ref{eqn:localphi} (although the distinction is conventionally clear from context \cite{Wagner2007}).

Coexisting bilayer phases correspond to points on the free-energy surface $f(\phi^\textrm{t}, \phi^\textrm{b})$ (Fig.~\ref{landscape}) that share a common tangent plane \cite{Williamson2014}.  The equilibrium coexistence for given overall composition $(\Phi^\textrm{t}, \Phi^\textrm{b})$ minimises the total free energy $F$ subject to constraints relating the bilayer phases' area fractions $\theta_n$ and compositions $\phi^\textrm{t(b)}_n$ to the overall compositions:\
\begin{align}\label{eqn:constraints}
\sum^\textrm{phases} \theta_n = 1~,
\sum^\textrm{phases} \theta_n \phi^\textrm{t}_n = \Phi^\textrm{t}~,
\sum^\textrm{phases} \theta_n \phi^\textrm{b}_n = \Phi^\textrm{b}~.
\end{align}
$F$ is proportional to the height of the tangent plane at the given $(\Phi^\textrm{t}, \Phi^\textrm{b})$. Coexistences not fully minimising $F$ are metastable.

\begin{figure*}[floatfix]
\includegraphics[width=17.0cm]{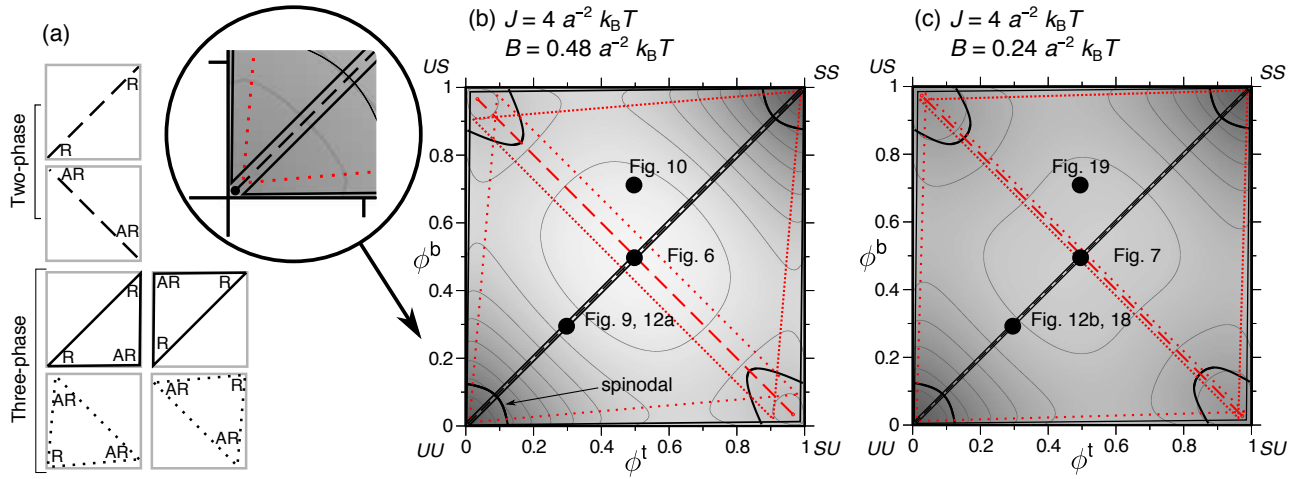}
\caption{\label{phasediag}(Color online) (a)~Shorthand diagrams for the metastable (AR-AR, AR-AR-R) and equilibrium (R-R, R-R-AR) coexistences studied in this work.
(b,c)~Grey contour plots of free-energy density $f(\phi^\textrm{t},\phi^\textrm{b})$ with partial phase diagrams overlaid. $\Delta_0 = 2\,a$, $\kappa = 3\,a^{-2}k_\textrm{B}T$, $V=0.6\,k_\textrm{B}T$, with $J$ and $B$ as indicated. Two-phase central R-R and AR-AR tie-lines (dashed) and three-phase triangles (solid or dotted) are shown. Metastable coexistences are red. A comprehensive version of (b) and detailed discussion appears in Appendix~\ref{app:details}. 
}
\end{figure*}

Partial phase diagrams for two of the main parameter points to be used in this work (except Figs.~\ref{0505J04B048}, \ref{messy} and Section~\ref{sec:weak}) appear in Fig.~\ref{phasediag}b,c. Spinodal lines enclose regions of local stability. An overall composition $(\Phi^\textrm{t}, \Phi^\textrm{b})$ within a tie-line (two-phase) or triangle (three-phase) can split into coexisting bilayer phases defined by the endpoints. The bilayer phases are, qualitatively, R (both leaflets dominated by the same species, approximate compositional symmetry between leaflets) or AR (opposite species, strong compositional asymmetry). R phases thus contain a majority of either $SS$ or $UU$ sites, and AR mostly $SU$ or $US$.
The precise composition of a given phase depends quantitatively on which phase coexistence it is part of. The separate leaflet compositions in a bilayer phase are given by the projections onto the $\phi^\textrm{t(b)}$ axes \cite{Collins2008, Wagner2007, Putzel2008}.
A state of phase coexistence is denoted R-R (two compositionally symmetric bilayer phases), R-R-AR (two symmetric phases and a strongly asymmetric one \cite{Collins2008}), AR-AR \cite{Perlmutter2011}, etc.
Overlapping tie-lines or triangles on the phase diagram imply that multiple phase coexistences are possible for a given $(\Phi^\textrm{t}, \Phi^\textrm{b})$ \cite{Williamson2014}.
 Fig.~\ref{phasediag}a establishes shorthand diagrams for the equilibrium and metastable coexistences appearing in Fig.~\ref{phasediag}b,c.  

In \cite{Galimzyanov2015} it is claimed that, for overall symmetric leaflets ($\Phi^\textrm{b}\!=\!\Phi^\textrm{t}$),  ``perfect antiregistration'' is possible only if $\Phi^\textrm{b}\!=\!\Phi^\textrm{t}\!=\!0.5$ (cf.\ Eq.~\ref{eqn:range}). It is true that AR-AR coexistence (strong local asymmetry throughout the bilayer) is possible only in a limited region centred on $\Phi^\textrm{b}\!=\!\Phi^\textrm{t}\!=\!0.5$ (see Appendix~\ref{app:details}). Similarly, conventional R-R coexistence is available for $\Phi^\textrm{b}\!=\!\Phi^\textrm{t}$, but may not be for $\Phi^\textrm{b}\!\neq\!\Phi^\textrm{t}$ (in \cite{Collins2008}, this lead to R-R-AR coexistence, corresponding to the black triangles on Fig.~\ref{phasediag}b,c).
However, even for overall leaflet compositions that prohibit one or both of ``perfect antiregistration'' (AR-AR) and conventional R-R coexistence, Fig.~\ref{phasediag} shows that there can still be \textit{multiple available phase coexistences}. The available states contain greater or lesser amounts of R versus AR phases and hence degrees of symmetry than the hypothetical case of fully independent leaflets.

For instance, AR-AR-R coexistence can be considered the ``minimally registered'' state where overall composition prohibits AR-AR coexistence;\ all bilayer regions are antiregistered except for one residual R phase. AR-AR-R may compete with the more registered R-R (for $\Phi^\textrm{b}\!=\!\Phi^\textrm{t}$) or R-R-AR \cite{Collins2008} (for $\Phi^\textrm{b}\!\neq\!\Phi^\textrm{t}$) states.
This behaviour is shown in the simulation results presented below. 

As explained in an upcoming Comment \cite{Comment}, the AR-AR-R state is relevant in evaluating the claim of \cite{Galimzyanov2015} that common observations of R-R can be explained via line energies alone, without a direct inter-leaflet coupling that would lead to an explanation from \textit{bulk} free energies. In the Comment \cite{Comment} we show that the elastic theory in \cite{Galimzyanov2015} implies that R-R \textit{may} be lower in total line energy than AR-AR-R for some compositions and parameters, but that for other compositions and parameters R-R is not equilibrium on the basis of line energies alone.

\color{black}

Fig.~\ref{phasediag}b,c are qualitatively similar;\ in both (as in the schematic landscape Fig.~\ref{landscape}) R minima are lower than AR so that equilibrium coexistence is R-R or R-R-AR (or R-AR, see Appendix~\ref{app:details}), depending on overall composition. A larger direct coupling $B$ (Fig.~\ref{phasediag}b) further raises the free energy of AR minima due to the energy cost of $SU$ and $US$ sites, hence the phase coexistence regions are quantitatively different between Fig.~\ref{phasediag}b,c.
Other parameter choices can yield AR minima lower than the R, so that AR-AR and AR-AR-R become equilibrium, not metastable, states. 

For simplicity, Fig.~\ref{phasediag} shows only the tie-lines and triangles studied in the present simulations. A comprehensive version of Fig.~\ref{phasediag}b, with further discussion of bilayer phase diagrams in the context of existing work and experimental data, appears in Appendix~\ref{app:details}. 

\subsection{Registration is nonconserved}

Since we assume no flip-flop or solvent exchange, the overall leaflet compositions $\Phi^\textrm{t(b)}$ are conserved \cite{Putzel2008}. However, depending on the phase coexistence the bilayer chooses, the overall number of pairwise registered lipids varies;\ AR ($SU$ or $US$) and R ($SS$ or $UU$) sites interconvert via
\begin{equation}\label{eqn:convert}
SU + US \rightleftarrows SS + UU~.\notag
\end{equation}
\noindent The \textit{nonconserved} overall occupancies $\mathcal{N}_{\alpha}$ of each site type $\alpha \in \{SS,UU,SU,US\}$ are constrained by the \textit{conserved} overall leaflet compositions
\begin{align}
\mathcal{N}_{SU} - \mathcal{N}_{US} &= (\Phi^\textrm{t} - \Phi^\textrm{b}) \mathcal{N}~,\\
\label{eqn:sc4}
\mathcal{N}_{SS} + \mathcal{N}_{SU} &= \Phi^\textrm{t} \mathcal{N}~,\\
\label{eqn:sc5}
\mathcal{N}_{UU}+\mathcal{N}_{US} &= (1-\Phi^\textrm{t}) \mathcal{N}~.
\end{align}
\noindent This leaves a free variable
\begin{equation}
\lambda \equiv \frac{\mathcal{N}_{SS} + \mathcal{N}_{UU}}{\mathcal{N}}~,
\end{equation}
\noindent the degree of molecule-level transbilayer symmetry -- the proportion of lipids, over the whole bilayer, that microscopically appose one of the same species. It can vary in the range
\begin{equation} \label{eqn:range}
\lvert \Phi^\textrm{t} +\Phi^\textrm{b} - 1\rvert \leq \lambda \leq 1 - \lvert \Phi^\textrm{t} - \Phi^\textrm{b} \rvert ~.
\end{equation}
Hence, as well as domain coarsening of a given coexistence of phases (during which $\lambda$ is roughly constant), the bilayer may switch between competing phase coexistences (Section~\ref{sec:bilphasdiag}), leading to a significant change in $\lambda$ while still conserving the overall leaflet compositions $\Phi^\textrm{t(b)}$. Therefore, $\lambda$ enables us to monitor such transitions. Its value $\lambda_0$ at $t = 0$ corresponds to the high-temperature limit:\
\begin{equation} \label{eqn:uncorr}
\lambda_0 = \Phi^\textrm{t} \Phi^\textrm{b} + (1- \Phi^\textrm{t})(1- \Phi^\textrm{b})~.
\end{equation}
\noindent $\lambda_0$ can be interpreted as the value that would be taken if the leaflets were completely independent of one another, coupled neither by a direct interaction or their combined hydrophobic thickness. 

Experimental measurement of $\lambda$ could be achieved using FRET (F\"{o}rster resonance energy transfer), with the donors and acceptors distributed exclusively in opposite leaflets. 

\subsection{Qualitative mapping onto conventional ternary phase diagrams}\label{sec:qual}

To properly account for transbilayer symmetry/asymmetry in describing phase behaviour, our model uses a separate composition axis for each leaflet \cite{Collins2008, Wagner2007, Putzel2008}. We now discuss how conventional bilayer phase diagrams map onto this description.

\begin{figure}[floatfix]
\includegraphics[width=8.0cm]{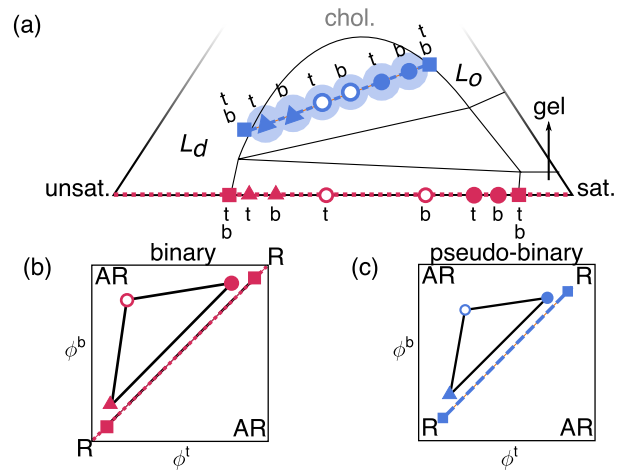}
\caption{\label{ternarybinary}(Color online) (a)~Schematic conventional ternary phase diagram. (b)~Leaflet compositions for coexistences in a binary (lipid-lipid) bilayer, using R-R and R-R-AR coexistences as examples. Pink symbols in (b) correspond to pink symbols in (a), and letters ``t'' and ``b'' in (a) label whether the symbol refers to top or bottom leaflets, or both. (c)~For pseudo-binary $L_o$-$L_d$ coexistence, the relationship to the ternary phase diagram is less precise. The corresponding blue symbols in (a) have halos indicating the uncertainty in ternary phase space.}
\end{figure}

Fig.~\ref{ternarybinary}a shows a typical lipid phase diagram of a ternary bilayer, which is taken to describe the fully registered case in which exact compositional symmetry holds throughout the bilayer. The binary cholesterol-free edge (sat.--unsat.) of Fig.~\ref{ternarybinary}a corresponds to the $\phi^\textrm{b}\!=\!\phi^\textrm{t}$ diagonal of Fig.~\ref{ternarybinary}b. The segment of that edge linking $L_d$ and gel corresponds to the R-R tie-line of Fig.~\ref{ternarybinary}b (squares). Because the leaflet compositions of R phases involved in R-R-AR coexistence will, in general, differ quantitatively from those involved in R-R coexistence, the mapping to a conventional ternary phase diagram is non-trivial.
R phases in the R-R-AR region (circles and triangles) of Fig.~\ref{ternarybinary}b do not lie precisely on the $\phi^\textrm{b}\!=\!\phi^\textrm{t}$ diagonal, so only have \textit{approximately} symmetric local leaflet compositions.
Thus, for the corresponding symbols on Fig.~\ref{ternarybinary}a the top and bottom leaflet compositions are slightly separated. The leaflet compositions for the strongly-asymmetric AR phase (open circle) lie towards opposite ends of the $L_d$-gel binodal.

Similarly, a binary model can represent a ternary bilayer in, e.g., an $L_o$-$L_d$ region of Fig.~\ref{ternarybinary}a.
This idea is exploited in simulation \cite{Sornbundit2014, Bagatolli2009} and theory \cite{Wagner2007, May2009, Putzel2008, Collins2008}, and a careful derivation by Garb\`{e}s Putzel and Schick \cite{Putzel2008} shows that it entails assuming the key compositional order parameter is the relative abundance of saturated versus unsaturated lipids. Two-component Ising universality in $L_o$-$L_d$ experiments \cite{Honerkamp2008} supports such a mapping. Fig.~\ref{ternarybinary}c illustrates this mapping for our bilayer phase diagram. The R-R tie-line of Fig.~\ref{ternarybinary}c (squares) corresponds to an $L_o$-$L_d$ tie-line of Fig.~\ref{ternarybinary}a, but the phases involved in the R-R-AR coexistence region need not map to leaflet compositions on precisely the same $L_o$-$L_d$ tie-line. Thus, the correspondence of a binary model to a pseudo-binary region of a real ternary system is approximate but the important qualitative features should be preserved.

\section{Simulation protocol and parameters}
 
\subsection{Protocol}\label{sec:protocol}

We simulate the model with a Kinetic Monte Carlo algorithm resembling Kawasaki (spin-exchange) dynamics \cite{Kawasaki1966}. The moves are:\ lateral exchanges of lipids within a single leaflet;\ coordinated lateral exchanges, where the top and bottom lipids of a lattice site move as one;\ and hydrophobic length changes of single lipids, proposed uniformly between $\pm 0.5\,a$. The total lattice size is $\mathcal{L}^2 = 200^2 = \mathcal{N}$. 
Time $t$ is measured Monte Carlo Steps (MCS), comprising $2\mathcal{N}$ lateral exchanges, $2\mathcal{N}$ combined lateral exchanges, and $2\mathcal{N}n_r$ length changes. We set $n_r = 10$, so that lipids relax their lengths faster than they laterally diffuse \cite{Wallace2006}. The bilayer is initialised with the lipids in each leaflet randomly distributed and the lengths initially relaxed $\sim1000$ times per lipid. 
The simulation models an instantaneous quench from high temperature at $t=0$. 

With idealised dynamics it is difficult to assign a corresponding physical time unit to the MCS. One could attempt to use characteristic self or collective diffusion times of the simulation, but a more straightforward comparison to experiment or molecular simulations could be made on the basis of \textit{domain size after a quench}, rather than \textit{time} after a quench. We can note that $a^2$ is the lateral area of one lipid. (Thus, for example, the crossover to registered domains occuring in Fig.~\ref{0505J4B048} is complete by the time the domain lengthscale reaches $\sim 100\,a$.) In \cite{Williamson2015a} we calculate nucleation energetics to estimate critical domain sizes for registration, presumably related to the lengthscale at which this crossover takes place. 

Similarly, hydrodynamics are not included, but will dominate the kinetics as domains exceed a certain length scale \cite{Siggia1979}, estimated by Fan~et~al.\ \cite{Fan2010} to be $\sim 10^{-6}\, \textrm{m}$ in lipid bilayers. However, hydrodynamics cannot affect the free-energy landscape, which determines the competing metastable and equilibrium states. Hence we expect the competition of registered and antiregistered phases studied here to remain robust. It is of great interest to consider marrying hydrodynamics to a coarse-grained bilayer free energy that accounts for both leaflets and their couplings, such as that derived from our model \cite{Williamson2014}.

\subsection{Registered and antiregistered structure factors}

We also monitor domain growth using separate structure factors for R and AR lattice sites $S_\textrm{m}(q,t)$ (where $\textrm{m} = \textrm{R},\textrm{AR}$). Their first moments \cite{Amar1988}
\begin{equation}
q_{\textrm{m}}(t)  \equiv \frac{ \sum_{q} q S_\textrm{m}({q},t)}{\sum_{q} S_\textrm{m}({q},t)}~,
\label{eqn:moments}
\end{equation}
\noindent lead to characteristic lengthscales for domains of microscopically registered versus antiregistered lipids
\begin{equation}
s_{\textrm{m}}(t)  \equiv \frac{2\pi}{q_{\textrm{m}}(t)}~.
\label{eqn:sizes}
\end{equation}
\noindent In the results presented here $s_\textrm{m}(t)$ is averaged over twenty independent trajectories. $S_\textrm{m}(q,t)$ can be related to more conventional structure factors suitable for experimental measurement (see Appendix~\ref{app:regantireg}).

\subsection{Parameters} \label{sec:params}

Detailed discussion of parameterisation is given in \cite{Williamson2014}. The lattice spacing is $a \sim 0.8\, \textrm{nm}$. We use $\kappa = 3\,a^{-2}k_\textrm{B}T$ ($\kappa_{A} \approx 60\,k_\textrm{B}T\textrm{nm}^{-2}$, in the range for lipid bilayers at $300\,\textrm{K}$ \cite{Wallace2005, Needham1990, Rawicz2000}). A fiducial estimate of the indirect coupling parameter is $J \sim 2\,a^{-2}k_\textrm{B}T$ \cite{Williamson2014}. $B$ leads to an effective value of $\gamma$ which has been widely estimated in the literature as $\gamma \sim 0.01 - 1\, k_\textrm{B}T\textrm{nm}^{-2}$ \cite{Pantano2011, Risselada2008, Polley2013, Putzel2011, May2009, Blosser2015}. In most simulations here we set the length/structural mismatch parameter $\Delta_0 = 2\,a \sim 1.6\,\textrm{nm}$. This relatively large value strengthens the indirect and direct coupling effects, making the kinetic regimes in the simulations clearer to interpret. Simulations with a smaller value $\Delta_0 \approx 1\,a$ yield similar kinetics (Section~\ref{sec:weak}), i.e., the balance of $J$ and $B$ determines what kinetics occurs.

In the theoretical plots (Figs.~\ref{phasediag} and \ref{stabdiag}) we use $V=0.6\,k_\textrm{B}T $ to exceed the threshold $V_0 \equiv 0.5\,k_\textrm{B}T$ required, in mean-field, for phase separation in the absence of any other couplings. However, the equivalent threshold in simulation (where fluctuations are included) is $V_0^\textrm{sim.} = 0.88\,k_\textrm{B}T$.  We employ $V=0.9\, k_\textrm{B}T$ for the simulations in an attempt to ensure a similar qualitative regime as the theoretical predictions. We have also tested the theory's sensitivity to $V$ by replotting Fig.~\ref{stabdiag} with $V=0.9\, k_\textrm{B}T$ (Fig.~\ref{stabdiagv09} in Appendix~\ref{app:linstab}), finding only a small difference in the predicted R/AR competition. This suggests the precise value of $V$ is not of crucial importance to our present study, so long as $V > V_0$ (mean-field) or $V > V_0^\textrm{sim.}$ (simulation).

\subsection{Flip-flop}

The present modelling does not include flip-flop. Wholesale redistribution of lipids between leaflets may influence phase behaviour \cite{Collins2008, Lin2006}. In \cite{Collins2008}, bilayers of asymmetric overall leaflet compositions $\Phi^\textrm{b} \neq \Phi^\textrm{t}$ appeared to relax over hours to overall symmetry $\Phi^\textrm{b}\!=\!\Phi^\textrm{t}$ through flip-flop. Such progression toward overall symmetry is perhaps the most intuitive effect of flip-flop, but in \cite{Lin2006}, in contrast, a slow decay towards an overall asymmetric state was inferred. While it seems that experiments on phase separation can be safely performed before flip-flop becomes important \cite{Collins2008, Visco2014}, we are currently extending our modelling to include the effects of flip-flop.
\section{Kinetic considerations}
\subsection{Initial instability and nucleation kinetics}
\label{sec:instnuc}
\begin{figure}[floatfix]
\includegraphics[width=8.0cm]{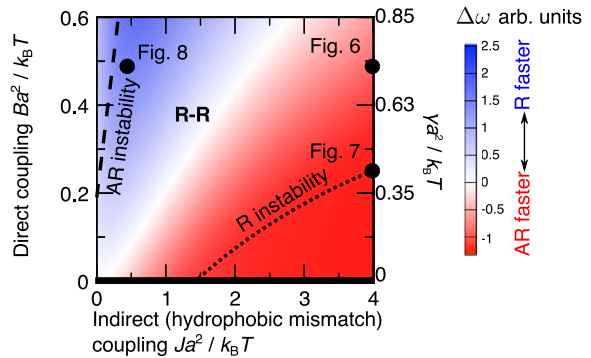}
\caption{\label{stabdiag}(Color online) Stability diagram for a bilayer of $(0.5,0.5)$ overall composition, with $\Delta_0 = 2\,a$, $\kappa = 3\,a^{-2}k_\textrm{B}T$, $V=0.6\,k_\textrm{B}T$, calculated from the mean-field free energy \cite{Williamson2014}. The equilibrium is R-R coexistence (R free-energy minima are lower than AR) except for a tiny region $Ba^2/k_\textrm{B}T \lesssim 0.005$ which we do not study here. Below the R instability line, the initial homogeneous state is not unstable to the R mode although R-R separation is still equilibrium \cite{Williamson2014}. Above the AR instability line, the AR minima do not exist (Appendix~\ref{app:details}) and there is no instability to the AR mode.
Colours show the relative growth rates of R and AR modes.}
\end{figure}

Consider $\Phi^\textrm{b}\!=\!\Phi^\textrm{t}\!=\!0.5$, which we label $(0.5,0.5)$. After quenching to an unstable region, the local composition is initially uniform at $(0.5,0.5)$.
Linear stability analysis then yields growth rates of symmetric (R) or asymmetric (AR) modes corresponding to demixing perturbations along the R-R or AR-AR central tie lines of Fig.~\ref{phasediag}b \cite{Williamson2014}, illustrated by the curved arrows on Fig.~\ref{landscape}. The analysis involves a Ginzburg-Landau free energy 
\begin{align} \label{eqn:fgrad}
F_\textrm{G-L} = \int d^2 r \left((f/a^2) + f_\textrm{grad} \right)~,
\end{align}
\noindent where $f_\textrm{grad} = \tfrac{1}{2}\tilde{J}(\nabla \bar{d})^2 + {V}(\nabla \phi^\textrm{t})^2 + {V} (\nabla \phi^\textrm{b})^2$.
The bulk term $f$ drives phase separation, while $f_\textrm{grad}$ penalises resultant gradients in composition and bilayer thickness. The balance determines which mode is faster and thus dominates initial phase separation (Fig.~\ref{stabdiag}). Further details can be found in Appendix~\ref{app:linstab} and \cite{Williamson2014}. 

For both R and AR modes, the Ising-like term (length-independent interactions, $V$) promotes instability via $f$ and penalises composition gradients via $f_\textrm{grad}$. The hydrophobic mismatch penalty $J$ plays a subtler role \cite{Williamson2014}. In $f_\textrm{grad}$ it penalises thickness gradients (only incurred by the R mode), without necessarily boosting R instability through the bulk term $f$. Hence, the AR mode can be faster although R-R coexistence is equilibrium. This is an explicit derivation of the physics underlying Ostwald's heuristic rule of stages (in this particular system), i.e., that a system will access metastable states on its way to the lowest free energy state.
Away from $(0.5,0.5)$ overall composition, competing instabilities will be more complex than the modes studied in Fig.~\ref{stabdiag}, but the competition between local symmetry and local asymmetry remains similar. Parameter ranges for typical phospholipids imply that these competing modes will be of comparable magnitude \cite{Williamson2014} so that, e.g., increasing the tail length mismatch could tip the balance to make the AR mode dominant.

Nucleation is required to reach equilibrium if initial demixing leads to a metastable state, although a bilayer of $(0.5,0.5)$ composition is conventionally thought to separate directly into registered phases by spinodal decomposition. In the cases studied here, the metastable coexistence is AR-AR or AR-AR-R, and equilibrium is R-R or R-R-AR, thus requiring nucleation of registered phases. In the nucleation process an area-dependent energetic benefit of converting AR to R phases competes with a cost from thickness mismatch at the R domain boundary;\ this is studied in \cite{Williamson2015a}.
The hydrophobic mismatch penalty $J$ penalises thickness mismatch at the boundary of a nucleating registered domain. Thus, as well as kinetically favouring AR-AR separation immediately after the quench, hydrophobic mismatch inhibits the nucleation required to equilibrate from a metastable state.
The direct coupling $B$ counters both these effects. It promotes initial instability to the R mode, and raises the AR free-energy minima relative to R (Fig.~\ref{phasediag}b versus c), so increasing the area-dependent benefit of nucleating the equilibrium R phases. 

Thus, depending on the competing inter-leaflet couplings, three classes of kinetics are possible \cite{Williamson2015a}:\ direct separation into equilibrium phases;\ nucleation out of a metastable state;\ and trapping in a metastable state. 

\begin{figure}[floatfix]
\includegraphics[width=8.0cm]{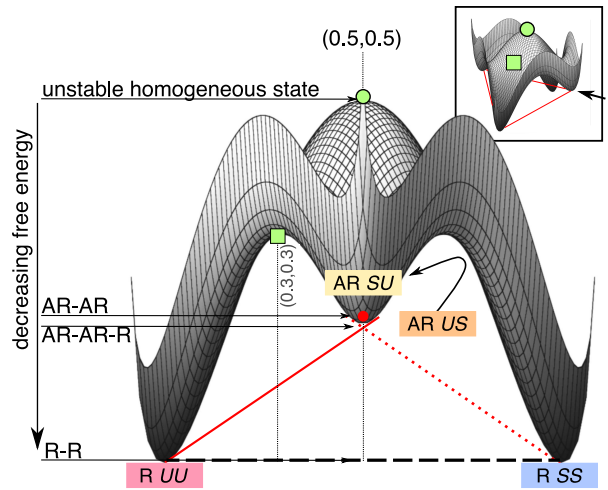}
\caption{\label{ARARR}(Color online) Schematic free-energy landscape of Fig.~\ref{landscape} viewed in the $(-1,1)$ direction (arrow, inset), such that AR minima appear one in front of the other. We superimpose tie-lines and triangles corresponding to the multiple coexistences available for $(0.5,0.5)$ overall composition:\ R-R equilibrium (black);\ two metastable AR-AR-R triangles (lines from this angle, red);\ metastable AR-AR (dot from this angle, red). One of the AR-AR-R triangles is illustrated on the inset.
}
\end{figure}

\subsection{Multiple metastable states}\label{sec:multiple}

A $(0.5,0.5)$ bilayer can lie within an R-R tie-line, two metastable AR-AR-R triangles and a metastable AR-AR tie-line, and thus access multiple metastable phase coexistences. As illustrated for the schematic free energy landscape in Fig.~\ref{ARARR}, the total bulk free energy at AR-AR-R coexistence -- given by the height of the tangent plane at $(0.5,0.5)$ -- is lower than for the AR-AR central tie-line. Thus, AR-AR coexistence can convert to AR-AR-R for a slight drop in bulk free energy (or, of course, reach equilibrium R-R to fully minimise free energy).
The AR-AR-R planes are degenerate for $(0.5,0.5)$, so a metastably trapped bilayer could locally fluctuate between $SS$ or $UU$ R phases in different regions (this seems to occur in Section~\ref{sec:weak} where weaker inter-leaflet couplings make fluctuations more apparent). 
 In contrast, an off-equimolar overall composition on the $\phi^\textrm{b}\!=\!\phi^\textrm{t}$ diagonal (e.g., $(0.3,0.3)$) can access a single AR-AR-R state, though equilibrium is still R-R. 
Appendix~\ref{app:details} contains a detailed diagram of all coexistence regions from the phase diagram in Fig.~\ref{phasediag}b, and discussion of their interpretation in relation to experiments. 

\section{Simulated phase-transition kinetics}

We now use simulation to investigate the kinetics of competing phase coexistences.
We vary the overall leaflet compositions $\Phi^\textrm{t(b)}$ to study different metastable/equilibrium coexistences, and vary the indirect (hydrophobic mismatch) coupling $J$ and direct coupling $B$ to explore different kinetic regimes which we expect to lie within the parameter ranges of typical phospholipids. Increasing $J$ could be physically realised by increasing the tail length of the saturated lipid species to increase hydrophobic mismatch \cite{Perlmutter2011}. Increasing $B$ might be achieved by increased difference in tail unsaturation between species, increasing their structural mismatch. Unless otherwise noted, the other simulation parameters are $\Delta_0 = 2\,a$, $\kappa = 3\,a^{-2}k_\textrm{B}T$, $V=0.9\,k_\textrm{B}T$. As well as visual inspection, phase-separation kinetics are monitored by the degree of microscopic transbilayer symmetry $\lambda$ (which could be experimentally accessed, e.g., by FRET).
Visualisation was performed with OVITO \cite{OVITO}. Videos corresponding to Figs.~\ref{0505J4B048}--\ref{0507J4B048} are available online \cite{SI}.

\subsection{$(\Phi^\textrm{t},\Phi^\textrm{b})\!=\!(0.5,0.5)$:\ R-R versus AR-AR}

\begin{figure}[floatfix]
\includegraphics[width=8.0cm]{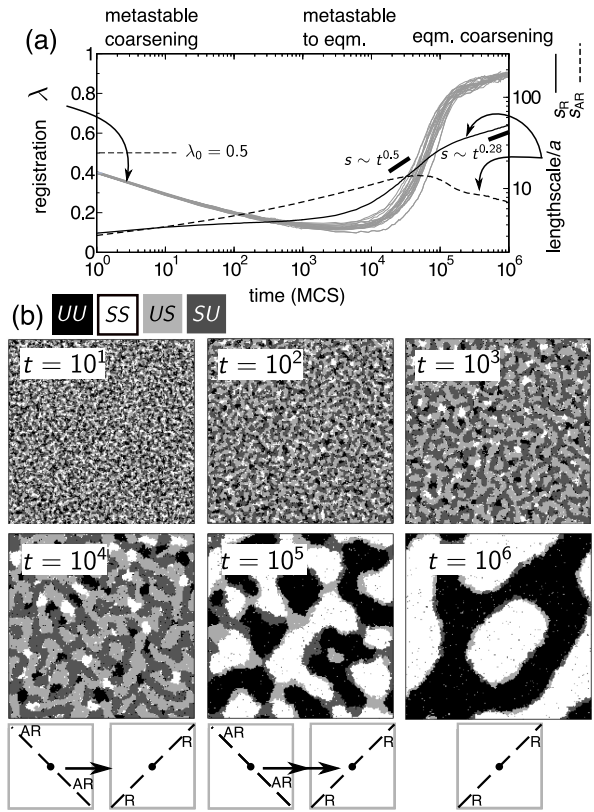}
\caption{\label{0505J4B048}(a) Grey:\ degree of pairwise lipid symmetry $\lambda$ through time (twenty independent runs) for a simulated bilayer of $(0.5,0.5)$ overall composition with $J=4\,a^{-2}k_\textrm{B}T$, $B=0.48\,a^{-2}k_\textrm{B}T$, $\Delta_0 = 2\,a$, $\kappa = 3\,a^{-2}k_\textrm{B}T$, $V=0.9\,k_\textrm{B}T$. The initial (uncorrelated leaflets) value is $\lambda_0 = 0.5$. Black (secondary axis):\ characteristic lengthscale of R and AR domains (averaged over the twenty runs). (b)~Trajectory snapshots showing the transition from AR-AR to R-R coexistence, causing nonmonotonic variation of $\lambda$. For the second row of snapshots, diagrams of the competing coexistences are shown, with overall composition marked by a dot. Videos corresponding to Figs.~\ref{0505J4B048}--\ref{0507J4B048} are available online \cite{SI}.}
\end{figure}

For $(0.5,0.5)$ overall composition, metastable AR-AR coexistence (local asymmetry throughout) competes with R-R (local symmetry throughout).
For $J=4\,a^{-2}k_\textrm{B}T$, $B=0.48\,a^{-2}k_\textrm{B}T$ (Fig.~\ref{0505J4B048}), we observe two-step kinetics. 
There is an immediate significant drop in $\lambda$ because $J > B$ favours pairwise antiregistration at the single-lipid level in the laterally homogeneous initial state \cite{Williamson2014} (in agreement with the experiments of \cite{Zhang2004}). Metastable AR domains form and coarsen (Fig.~\ref{0505J4B048}b). The bilayer reduces the hydrophobic cost of remaining R sites through annihilation into further AR sites, and coalescence into R domains.
These coalesced groups become nuclei of $SS$ and $UU$ phases \cite{Williamson2015a}. From $t \sim 10^4 - 10^5$, these nuclei grow at the expense of the AR phases, increasing $\lambda$ and converting the bilayer from metastable AR-AR to equilibrium R-R coexistence for a payoff in bulk free energy.
The stochastic nature of the nucleation transition is apparent by the spread in $\lambda$ between independent simulation trajectories during this period. Thereafter, coarsening of the equilibrium phases continues and $\lambda$ is roughly constant. 

Hence, the transition between the competing phase coexistences is signified by nonmonotonic evolution of $\lambda$. The bilayer first becomes more asymmetric than the uncorrelated case ($\lambda = \lambda_0$) as it favours AR-AR coexistence, then more symmetric as it switches to equilibrium R-R.
The two-step kinetics in Fig.~\ref{0505J4B048} is in agreement with Fig.~\ref{stabdiag};\ $J$ and $B$ for this parameter point are well within the region where the mean-field theory predicts the AR mode to dominate so that AR-AR occurs first, as in Fig.~\ref{0505J4B048}.

\begin{figure}[floatfix]
\includegraphics[width=8.0cm]{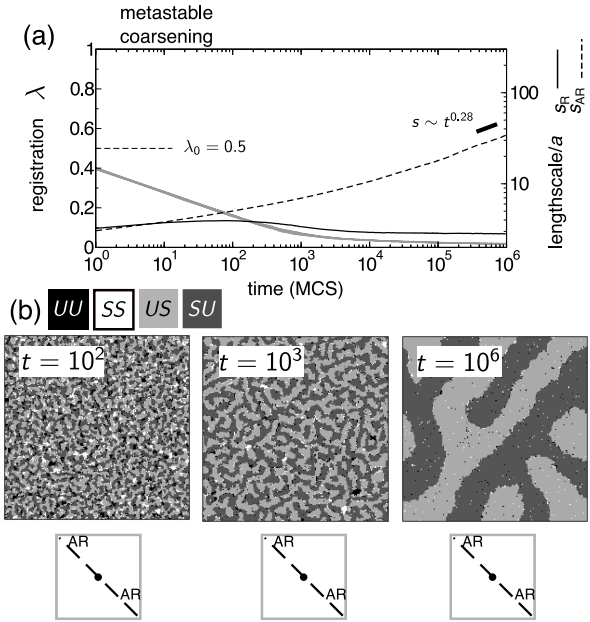}
\caption{\label{0505J4B024}$J=4\,a^{-2}k_\textrm{B}T$, $B=0.24\,a^{-2}k_\textrm{B}T$, $(0.5,0.5)$ overall composition. As Fig.~\ref{0505J4B048}, with weaker direct coupling $B$. Nucleation of R domains is suppressed and the bilayer is trapped in AR-AR coexistence.}
\end{figure}

\begin{figure}[floatfix]
\includegraphics[width=8.0cm]{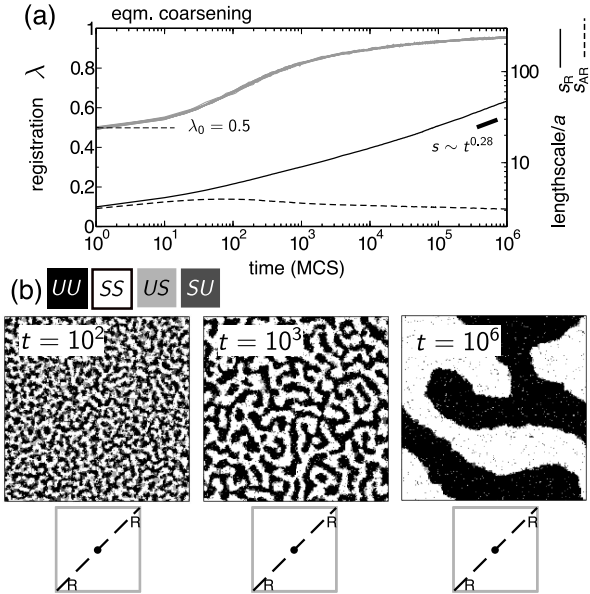}
\caption{\label{0505J04B048} $J=0.4\,a^{-2}k_\textrm{B}T$, $B=0.48\,a^{-2}k_\textrm{B}T$, $(0.5,0.5)$ overall composition.
As Fig.~\ref{0505J4B048}, with weaker indirect coupling $J$. The R mode is fastest, and the bilayer immediately accesses R-R coexistence.}
\end{figure}

The lengthscales $s_\textrm{R}$ and $s_\textrm{AR}$ show initially faster growth of AR domains. During conversion of AR to R domains, there is an increase in the growth exponent of registered domains to $\alpha_\textrm{R} \approx 0.5$ (where $s_\textrm{R} \sim t^{\alpha_\textrm{R}}$), while AR domain growth slows. This exponent is expected for the nonconserved Ising (Model A) universality class \cite{Vinals1988}, and we attribute it to nonconserved growth of R nuclei via annihilation of AR domains. As $\lambda$ begins to saturate, further conversion of AR slows.
Thus, $SS$ and $UU$ now play the role of ``up'' and ``down'' spins in the \textit{conserved} Ising model. 
R domain growth slows to $\alpha_\textrm{R} \approx 0.28$, a typical apparent exponent for diffusive coarsening in the conserved Ising (Model B) class \cite{Amar1988}, which would tend to $\alpha = \tfrac{1}{3}$ in the asymptotic time limit. We employ $\alpha \approx 0.28$ as a heuristic test for tending towards the conserved growth scaling regime. As domains grow larger these exponents will be altered by hydrodynamics (Section~\ref{sec:protocol}), which is not included in our model.

With weaker direct coupling $B=0.24\,a^{-2}k_\textrm{B}T$ (Fig.~\ref{0505J4B024}) the bilayer gets trapped in AR-AR coexistence and all R nuclei are destroyed, leaving the metastable phases to coarsen indefinitely.
AR domains coarsen with an exponent $\alpha_\textrm{AR} \approx 0.28$ reflecting conserved growth, $SU$ and $US$ now playing the role of up and down spins in the conserved Ising model. This metastably trapped AR-AR state resembles that found in molecular simulations with strong hydrophobic length mismatch \cite{Perlmutter2011, Reigada2015}, although in \cite{Perlmutter2011, Reigada2015} it could be stabilised due to finite simulation size \cite{Williamson2015a}.

For $B=0.48\,a^{-2}k_\textrm{B}T$ and weaker hydrophobic mismatch parameter $J=0.4\,a^{-2}k_\textrm{B}T$ (Fig.~\ref{0505J04B048}) the bilayer proceeds straight to R-R coexistence. The proportion of registered lipids $\lambda$ increases immediately, in contrast to the two-step behaviour seen in Fig.~\ref{0505J4B048}. R domains reach a growth exponent  $\alpha_\textrm{R} \approx 0.28$ without a regime of nonconserved ($\alpha_\textrm{R} \approx 0.5$) growth. 
This behaviour is consistent with the prediction of Fig.~\ref{stabdiag} that for these parameters the R mode is faster during initial phase separation.

\begin{figure}[floatfix]
\includegraphics[width=8.0cm]{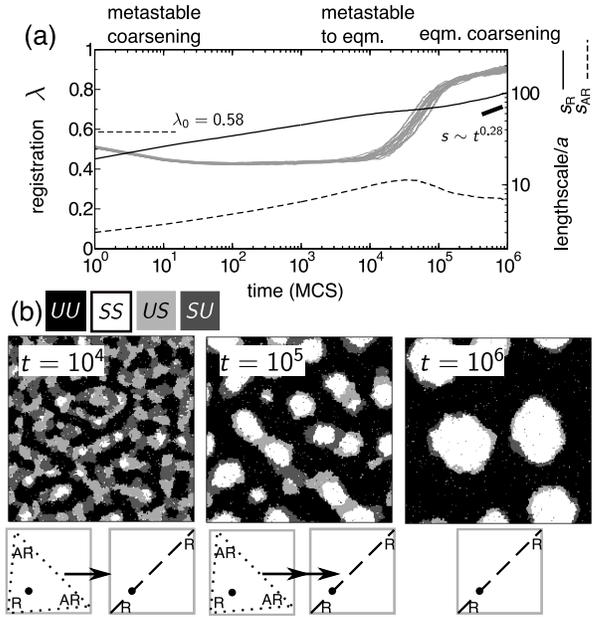}
\caption{\label{0303J4B048}$J=4\,a^{-2}k_\textrm{B}T$, $B=0.48\,a^{-2}k_\textrm{B}T$, $(0.3,0.3)$ overall composition.
As Fig.~\ref{0505J4B048}, for $(0.3,0.3)$ overall composition. The bilayer moves from AR-AR-R to R-R coexistence.}
\end{figure}

\begin{figure}[floatfix]
\includegraphics[width=8.0cm]{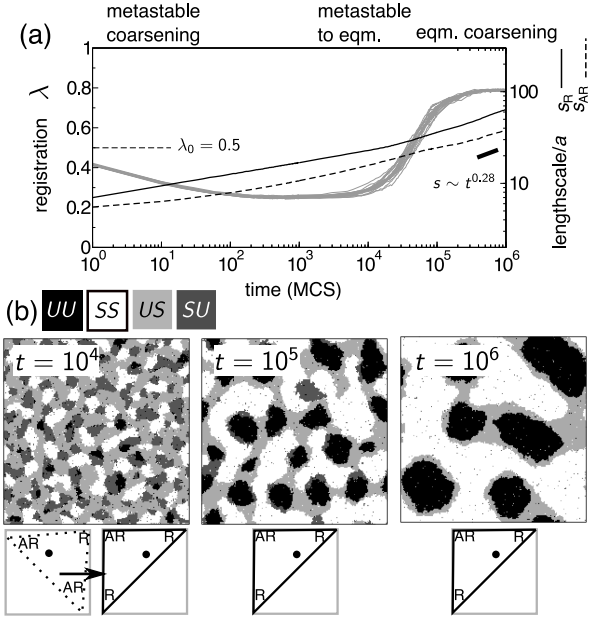}
\caption{\label{0507J4B048}$J=4\,a^{-2}k_\textrm{B}T$, $B=0.48\,a^{-2}k_\textrm{B}T$, $(0.5,0.7)$ overall composition.
As Fig.~\ref{0505J4B048}, for $(0.5,0.7)$ overall composition. The bilayer moves from AR-AR-R to R-R-AR coexistence.}
\end{figure}

\subsection{$(0.3,0.3)$:\ R-R versus AR-AR-R}

A common experimental case is leaflets containing the same, but nonequimolar mixture.
If the composition is within the R-R central tie-line, equilibrium coexistence is R-R. We use $\Phi^\textrm{b}\!=\Phi^\textrm{t}\!=\!0.3$, entailing excess $UU$ sites relative to the $(0.5,0.5)$ case. According to Figs.~\ref{phasediag}b,c, it is also within a metastable AR-AR-R coexistence region. 

Fig.~\ref{0303J4B048} shows the associated two-step kinetics. It is analogous to that for $(0.5,0.5)$ composition in Fig.~\ref{0505J4B048}, except that now the metastable state comprises two AR phases plus a $UU$ R phase. Again, $\lambda$ initially drops from the uncorrelated case $\lambda_0$, this time indicating a preference for the AR-AR-R state, then increases, as R-R is reached. Unlike in Fig.~\ref{0505J4B048}, there is no clear $\alpha_\textrm{R} \approx 0.5$ growth during AR to R conversion. This probably arises because the metastable state now contains AR-AR-R, not just AR-AR, so the transition to equilibrium coexistence involves not only AR to R conversion but also continued coarsening of the existing $UU$ domains. This is also shown in the behaviour of $\lambda$ which, in comparison to Fig.~\ref{0505J4B048}, does not attain as low a minimum, and so grows less dramatically. With smaller $B$ (Appendix~\ref{app:traj}), nucleation of the registered $SS$ phase is suppressed, leaving metastable AR-AR-R coexistence. 

These simulations thus demonstrate an important consequence of the discussion in Section~\ref{sec:bilphasdiag}:\ even where \textit{perfect} antiregistration (AR-AR) is not possible, ``minimally registered'' AR-AR-R is possible. This is in contradiction to \cite{Galimzyanov2015}, which uses the fact that AR-AR is only possible in a narrow window of compositions (Fig.~\ref{complete}b) to argue that R-R is generally stabilised by line energies alone, so that explaining R-R would not require direct inter-leaflet coupling (a \textit{bulk} free energy effect). In an upcoming Comment \cite{Comment} we show that R-R is stabilised by line energies only for certain compositions and parameters, because one must compare it to the AR-AR-R state if AR-AR is not available \footnote{The line energy of R-R can be less than AR-AR-R in a certain range of compositions \textit{only if} the line tension of R-R interfaces exceeds that of R-AR interfaces by less than a geometric factor determined by tie-lines in the phase diagram. According to the calculations in \cite{Galimzyanov2015} this occurs for some parameters but not others. The details will be given in \cite{Comment}.}.

\color{black}

\subsection{$(0.5,0.7)$:\ R-R-AR versus AR-AR-R}

We now consider asymmetric overall compositions, as in experiments on bilayers specially prepared with differing leaflets \cite{Collins2008}. Overall composition $(0.5,0.7)$, according to Figs.~\ref{phasediag}b,c, gives R-R-AR coexistence at equilibrium, as was observed in \cite{Collins2008}. Just as $(0.3,0.3)$ prevented perfectly asymmetric AR-AR, here $(0.5,0.7)$ prevents perfectly symmetric R-R. 

This R-R-AR state competes with AR-AR-R coexistence. In Fig.~\ref{0507J4B048}, two-step kinetics can be seen. We effectively have coarsening of a 2D ternary mixture ($SU$, $US$ and $UU$-dominated phases), as recently studied in \cite{Awaneesh2015}. Accordingly, we expect the same scaling as for a binary mixture and, indeed, both measures of domain size exhibit $\alpha_\textrm{R} \approx \alpha_\textrm{AR} \approx 0.28$ growth.
With smaller $B$ (Appendix~\ref{app:traj}), the bilayer is metastably trapped in AR-AR-R. If the overall leaflet compositions are strongly asymmetric, the equilibrium state may be R-AR instead of R-R-AR. This is discussed in Appendix~\ref{app:details} in relation to some available experimental findings \cite{Lin2015, Visco2014, Collins2008}.

\begin{figure}[floatfix]
\includegraphics[width=8.0cm]{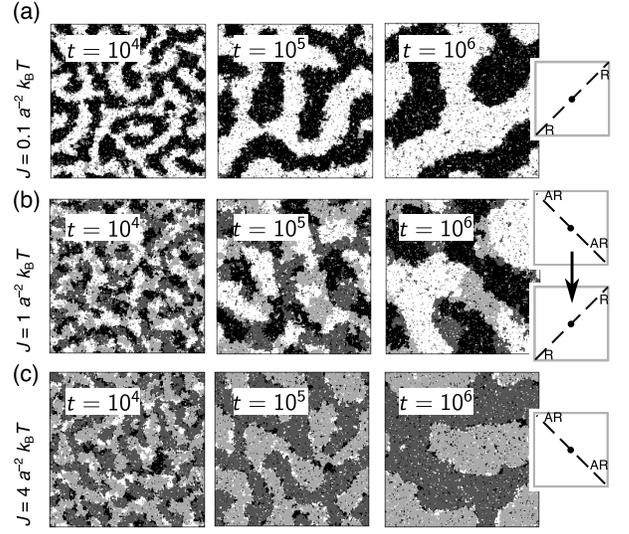}
\caption{\label{messy}Trajectories using $\Delta_0 = 1\,a$, which weakens both indirect and direct couplings \cite{Williamson2014}. The overall composition is $(0.5,0.5)$.
$B=0.3\,a^{-2}k_\textrm{B}T$, and $J$ is increased. 
(a)~$J=0.1\,a^{-2}k_\textrm{B}T$. Immediate separation into R-R coexistence (cf.\ Fig.~\ref{0505J04B048}). 
(b)~$J=1\,a^{-2}k_\textrm{B}T$. Competing AR-AR and R-R coexistences, slowly converting to R-R domination (cf.~Fig.~\ref{0505J4B048}). 
(c)~$J=4\,a^{-2}k_\textrm{B}T$. Metastably trapped in AR-AR (cf.\ Fig.~\ref{0505J4B024}).}
\end{figure}

\begin{figure*}[floatfix]
\includegraphics[width=16.5cm]{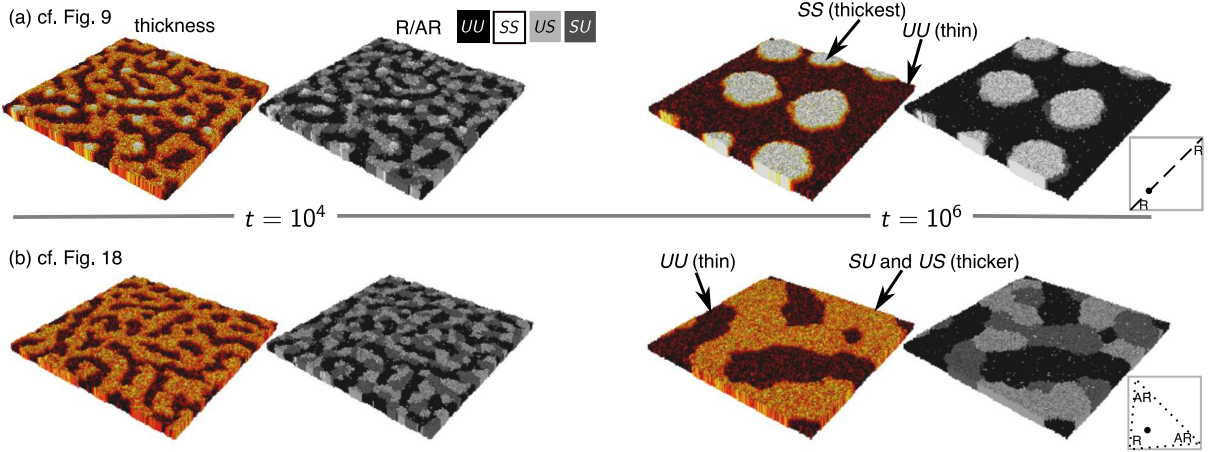}
\caption{\label{3D}(Color online) (a)~Snapshots from the trajectory from Fig.~\ref{0303J4B048} rendered in pseudo-3D. Colours showing total bilayer thickness as would be measured via AFM (analogous to total bilayer fluorescence in standard fluorescence microscopy) are compared with greyscale showing leaflet registration and antiregistration (as in Fig.~\ref{0303J4B048}).
(b)~Snapshots from a trajectory with weaker direct inter-leaflet coupling $B$ (see Appendix~\ref{app:traj}). Nucleation of $SS$ is suppressed, leaving metastable three-phase coexistence which could be mistakenly interpreted as two-phase equilibrium in the thickness representation.}
\end{figure*}

\subsection{Weaker mismatch}\label{sec:weak}

So far we have used a mismatch parameter $\Delta_0 = 2\,a \approx 1.6\,\textrm{nm}$ to strengthen the indirect and direct couplings and more easily distinguish the kinetic regimes. With a smaller value $\Delta_0 = 1\,a$, both the indirect and direct couplings become weaker, but estimated phospholipid parameters still imply competing R and AR modes of comparable magnitude \cite{Williamson2014}.
Fig.~\ref{messy} shows the same phenomenology with $\Delta_0 = 1\,a$ as in the previous simulations. As expected, increasing indirect coupling $J$ in Fig.~\ref{messy} has a similar effect to decreasing $B$:\
for small $J$ the R mode is fastest;\ for intermediate $J$, AR and R compete;\ for larger $J$, metastable trapping in AR-AR coexistence is evident as the energy cost for thickness mismatch prevents nucleation of R phases. In Fig.~\ref{messy}c, an AR-AR trapped state appears to locally fluctuate into AR-AR-R, creating transient regions of R phase. This is consistent with the discussion in Section~\ref{sec:multiple}.

\section{Imaging that does not resolve distinct leaflets}

Phase-separating bilayers are often imaged with the fluorescent tag distributed throughout both leaflets. In this method, one does not measure the separate leaflets, but rather the \textit{sum} of the two leaflets' compositions. This is analogous to measuring bilayer thickness via AFM (atomic force microscopy), by which the sum of the two leaflet thickness is measured --
coexisting AR phases ($SU$ and $US$) would be of the same thickness and thus not distinguishable from one another. We imitate this by imaging the simulated bilayer in a total thickness representation (Fig.~\ref{3D}), no longer labelling $SS$, $SU$ etc. The bilayer is rendered in pseudo-3D by assigning total thickness to $z$. (The reference $d_0 \equiv \ell_{S0} + \ell_{U0}$ is irrelevant to the behaviour of the model, but for Fig.~\ref{3D} we have set $d_0 = 5\,a \sim 4\,\textrm{nm}$.)

Fig.~\ref{3D}a shows snapshots corresponding to Fig.~\ref{0303J4B048}. During the transition from metastable to equilibrium coexistence, three distinct thicknesses, arising from \textit{four} distinct phases, appear. The bilayer initially prefers metastable AR-AR-R coexistence, where the R phase is $UU$. The two AR phases have identical, intermediate thickness, $UU$ is thinner and small nuclei of $SS$ are thickest.
The $SS$ and $UU$ phases then grow at the expense of AR ($SU$ and $US$) phases, leaving the two thicknesses of the $SS$ and $UU$ phases. The appearance of three distinct thicknesses and nonconservation through time of their respective area fractions would thus signify a transition between competing coexistences in molecular simulation or experiment. AFM would be well-suited, so that these kinetics could be monitored while domains are too small to resolve optically. 

In Fig.~\ref{3D}b the bilayer remains trapped in metastable AR-AR-R coexistence. In the absence of an $SS$ phase for comparison (or \textit{a priori} knowledge of its expected thickness) it is in principle possible to misinterpret the two AR phases as a single, registered phase. Hence, long-lived or steady state (i.e., trapped) metastable three-phase coexistence could masquerade as two-phase (R-R) equilibrium coexistence. The apparent area fractions of thicker and thinner phases differ from the true R-R equilibrium state.

However, studies in which leaflets are independently fluorescently labelled suggest the expected equilibrium of R-R-AR \cite{Collins2008} or R-R coexistence \cite{Garg2007} is reached, implying that such metastable trapping is avoided. Therefore, it is likely that experiments with similar model phospholipids also reach equilibrium phase coexistence, and the misinterpretation shown in Fig.~\ref{3D}b, requiring \textit{long-lived} metastability that is apparently equilibrium, does not arise. However, experiments limited by optical resolution would not probe the kinetics of a metastable to equilibrium transition on time/lengthscales before domains reach resolvable size. Large hydrophobic mismatch would increase the possibility of long-lived metastable AR-AR as seen in simulation \cite{Perlmutter2011, Stevens2005}, while Lin et al.\ \cite{Lin2006} found slow creation of AR domains at the expense of R, in a liquid-gel bilayer (perhaps influenced by the solid substrate). 

Labelling the leaflets separately provides extra insight by revealing whether the pattern of enrichment and depletion of dye in one leaflet is colocalised with that in the other. However, such colocalisation need not necessarily imply R-R coexistence (local compositional symmetry everywhere), but can also indicate R-AR (see Appendix~\ref{app:details} and \cite{Williamson2015a}).

\section{Discussion}

Evidence in the literature implies that the leaflets of a phase-separating bilayer are inherently coupled, both by a direct coupling ($B$) that encourages compositionally symmetric bilayer phases (registered, R), and an indirect coupling ($J$) from hydrophobic mismatch that favours strong compositional asymmetry (antiregistered phases, AR) \cite{Pantano2011, Risselada2008, Polley2013, Perlmutter2011, Stevens2005, Zhang2004, Zhang2007, Lin2006}. Here we have studied the resulting kinetics, by simulating the lattice model underlying the theory in \cite{Williamson2014}. We have found signatures by which the kinetics could be detected in molecular simulation or experiment, to provide insight into fundamental interactions.

Governed by the inter-leaflet couplings and the overall leaflet compositions, multiple phase coexistences involving different degrees of transbilayer symmetry can compete. An important result of the present work is that this applies over wide regions of the phase diagram. For instance, AR-AR coexistence (i.e., strong local asymmetry everywhere throughout the bilayer, which has been called ``perfect antiregistration'' \cite{Galimzyanov2015}), may only be possible in a small window of overall compositions (see Appendix~\ref{app:details}). However, outside this window, ``minimally registered'' AR-AR-R coexistence can still occur, as well as the typically-observed R-R \cite{Korlach1999} or R-R-AR \cite{Collins2008} states \footnote{We note in passing that a state of fully symmetric domains with very small asymmetric ``slip regions'' around the edges \cite{Galimzyanov2015} corresponds to R-R coexistence -- such slip regions do not constitute their own bulk phase.}. Hence, for a variety of overall compositions, a phase separating bilayer can choose between states that are either \textit{more or less} symmetric than the hypothetical ``uncorrelated'' case of completely independent leaflets (Eq.~\ref{eqn:uncorr}).

The direct inter-leaflet coupling interaction $B$ typically renders the more-asymmetric (AR-AR or AR-AR-R) states \textit{metastable}, due to their higher bulk free energy, providing an appealing explanation for observations of domain registration. Conversely, observed domain antiregistration \cite{Perlmutter2011, Reigada2015} may be understood as metastably trapped due to strong hydrophobic mismatch, or stabilised by small domain size \cite{Williamson2015a}. It was argued recently that domain registration observations are explained in general by line energies alone, \textit{without} direct inter-leaflet coupling \cite{Galimzyanov2015} -- our reasons for disputing this claim are detailed in an upcoming Comment \cite{Comment}.

Physical parameter estimates \cite{Williamson2014} imply that competing instabilities to form registered versus antiregistered domains can be of comparable strength, so that changes to molecule tail length or structure mismatch -- effectively tuning $J$ and $B$ -- may determine which occurs first after a quench. Subsequent to reaching a metastable state, the bilayer must nucleate to reach equilibrium \cite{Williamson2015a}. We also find a preference for pairwise lipid asymmetry in the initial homogeneous state, in line with \cite{Zhang2004, Zhang2007}. Hence, including both direct inter-leaflet coupling and hydrophobic mismatch effects helps unify observations of transbilayer symmetry \cite{Korlach1999, Dietrich2001,Collins2008} and asymmetry \cite{Perlmutter2011, Stevens2005, Zhang2004, Zhang2007, Lin2006} in mixed bilayers. The competing inter-leaflet couplings also lead to a ``critical radius'' for domain registration, below which AR domains are \textit{stable} \cite{Williamson2015a};\ this could be important in cellular rafts or clusters, which are thought to exist in a size range $10-100\,\textrm{nm}$ \cite{Lingwood2010} \footnote{It is important to note that a critical radius does not imply that antiregistered domains must \textit{automatically} switch to registered once they exceed a certain size \cite{Williamson2015a}.}.

The findings here can be probed by studying the kinetics of transbilayer symmetry upon quenching from high temperature. Conducting such studies either in molecular simulations or experiment can, via our theory, directly reveal the relative importance of the direct versus indirect inter-leaflet couplings \cite{Williamson2014}. Therefore, although the simulation used here is idealised (Section~\ref{sec:protocol}), it has the benefit of directly linking observed kinetics to a free-energy landscape derived from the same microscopic model;\ this is turn provides a basis for more realistic simulations and experiments to test the essential features captured in the model.
An interesting experiment would be a FRET study in which donors and acceptors are distributed in opposite leaflets, thus accessing the degree of microscopic transbilayer symmetry $\lambda$. A nonmonotonic evolution of the FRET signal after a rapid quench to a phase-separating region would be a clear signature of competing asymmetric and symmetric modes of phase separation.

\appendix

\section{Bilayer phase diagram details}\label{app:details}

\begin{figure*}[floatfix]
\includegraphics[width=17cm]{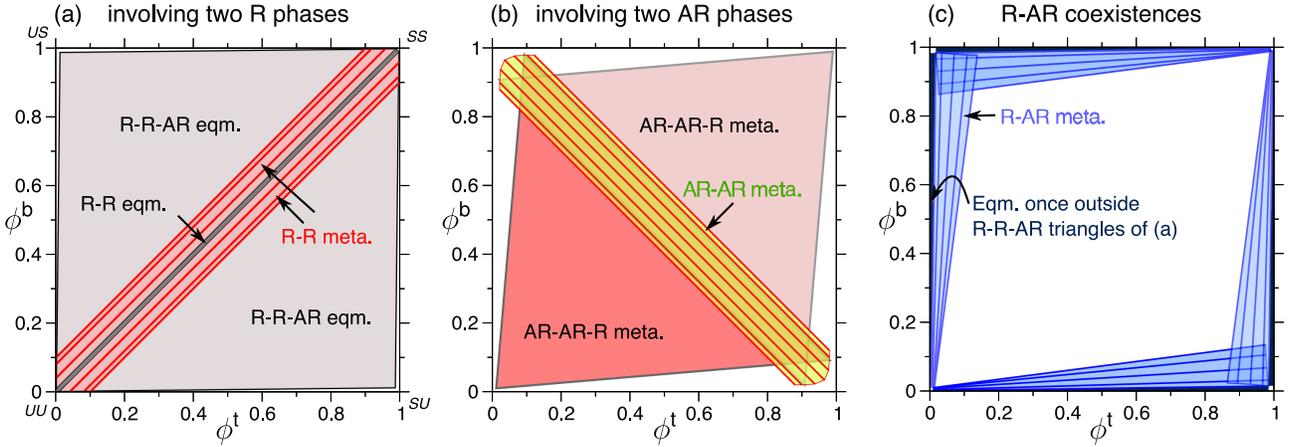}
\caption{\label{complete}(Color online) Phase-coexistence regions corresponding to the partial phase diagram shown in Fig.~\ref{phasediag}b. Coexistences are divided into those containing two R phases, two AR phases, or R-AR coexistence. (a)~A narrow two-phase R-R equilibrium (dark grey) region abuts two equilibrium three-phase R-R-AR regions (grey). A metastable continuation (red) extends each side of the equilibrium R-R region, and example tie-lines are plotted. (b)~Mutually overlapping three-phase metastable AR-AR-R triangles (reds), and a metastable two-phase AR-AR region (yellow) in which example tie-lines are plotted. (c)~Two-phase R-AR metastable (blue) and equilibrium (dark blue) regions. The equilibrium region obtains only outside the R-R-AR triangles of (a). With the present parameters the equilibrium R-AR regions are close to the edges where numerical evaluation is difficult (contrast Ref.~\cite{Williamson2014}), so they are sketched here by hand (dark blue).}
\end{figure*}

\begin{figure}[floatfix]
\includegraphics[width=7.5cm]{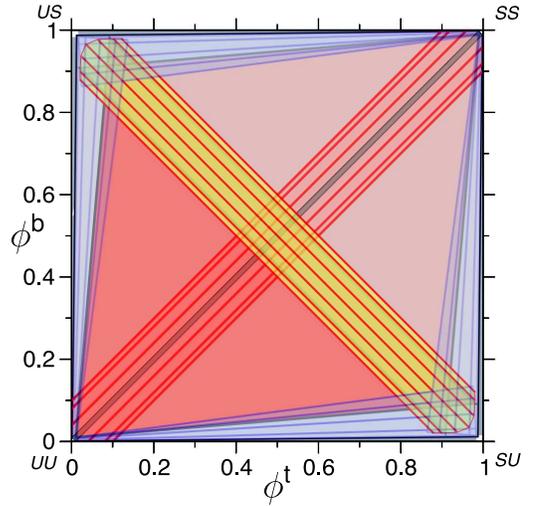}
\caption{\label{overlay}(Color online) Fig.~\ref{complete}a,b and c overlaid to create a full phase diagram.}
\end{figure}

Fig.~\ref{complete} provides detail on the phase diagram from Fig.~\ref{phasediag}b by explicitly demarcating all regions of phase coexistence, rather than isolated tie-lines and three-phase triangles. Fig.~\ref{overlay} shows all the phase-coexistence regions on a single diagram. We classify the coexistences into those with two R phases (Fig.~\ref{complete}a), two AR phases (Fig.~\ref{complete}b), or R-AR (Fig.~\ref{complete}c). As discussed in \cite{Williamson2014}, each two-phase equilibrium region is associated with a metastable promontory, its continuation into a region whose equilibrium is three-phase. Hence, the equilibrium R-R region of Fig.~\ref{complete}a is surrounded by a metastable R-R region, in which the equilibrium state is R-R-AR. In Fig.~\ref{complete}b, an AR-AR region overlaps two AR-AR-R triangles, which are mutually overlapping (the idea of multiple metastable coexistences for a given state point is shown in Fig.~\ref{ARARR}). All coexistences appearing in Fig.~\ref{complete}b are metastable because AR minima are higher in free energy than R. For certain parameter choices, this can be reversed \cite{Williamson2014}. 

R-AR coexistence is available for highly asymmetric overall compositions (Fig.~\ref{complete}c) but has not been the focus here. Equilibrium R-AR forms a continuation of the R-R-AR equilibrium triangles of Fig.~\ref{complete}a. Due to strong segregation for the present parameters, R-R-AR triangles are large;\ hence, equilibrium R-AR regions exist only very close to the edges of the phase diagram. Compare \cite{Williamson2014} in which larger equilibrium R-AR ``arms'' exist. .

The finite positive tilt of R-AR tie-lines implies that, if each leaflet were imaged \textit{separately}, domains of larger-than-average $\phi^\textrm{t(b)}$ appear in both leaflets and be spatially colocalised, although the overall leaflet compositions remain highly asymmetric \cite{Visco2014}.  Thus, for an asymmetric bilayer in R-AR coexistence, whether compositional domains in one leaflet ``induce'' them in the other is related to whether the R-AR tie-line is tilted \textit{enough} for them to be detected -- a fully horizontal or vertical R-AR tie-line would imply compositional domains in only one leaflet. This is discussed also in \cite{Williamson2015a}.

%For highly asymmetric overall compositions, R-AR may be the \textit{only} available phase coexistence. 
The tilt of R-AR tie-lines is a result of the direct inter-leaflet coupling $B$. 
Without $B$, the R and AR free-energy minima would be the same height \footnote{In fact, for $B=0$ the microscopic effects of hydrophobic mismatch make AR free-energy minima very slightly \textit{lower} than R, but that is not of concern here \cite{Williamson2014}.}, and all R-AR tie-lines would be vertical or horizontal in $(\phi^\textrm{t},\phi^\textrm{b})$ space. 
Thus, the same direct inter-leaflet coupling that can make R phases lower in free energy than AR (rendering R-R coexistence equilibrium for \textit{symmetric} overall leaflet compositions), also gives the ability of one leaflet to ``induce'' domains in the other leaflet where the overall compositions are strongly asymmetric \cite{Visco2014, Lin2015}. Further, R-AR tie-line tilt helps explain aspects of the R-R-AR coexistence observed \cite{Collins2008}. Such observations \cite{Visco2014, Lin2015, Collins2008} therefore support a direct inter-leaflet coupling interaction. 

In principle one could prepare a compositionally-pure bottom leaflet (for example) and a mixed top leaflet. In such a case, the top leaflet can form compositional domains while the bottom leaflet comprises domains of differing \textit{thicknesses} (implicitly, tail ordering) despite its uniform composition. Via the direct inter-leaflet coupling $B$, the thickness of the bottom leaflet is coupled not only to its own composition but to that of the top leaflet.
(Indeed, a one-component leaflet or a one-component \textit{bilayer} can separate into domains of qualitatively differing tail order, such as fluid and gel. However, the calculations in this paper are explicitly assumed to be far from such first-order transitions of the pure lipids -- see Footnote [25]).

The free-energy landscape and available phase equilibria (``phase diagram topology'') can qualitatively change for different parameter choices \cite{Wagner2007, Putzel2008}.
For example, strong enough $B$ causes AR minima to disappear entirely, in which case R-R-AR may still be possible, but AR-AR or AR-AR-R coexistence is not. If AR minima exist but are of sufficiently different height and curvature to the R minima, the AR-AR-R common tangent might disappear while AR-AR remains. In Fig.~\ref{complete}, the fourfold symmetric R-AR arms extend arbitrarily close to the edges of the phase diagram, but they could also become truncated \cite{Wagner2007} such that one pure leaflet (e.g., $\Phi^\textrm{b} \to 0$) causes the bilayer to lie outside any coexistence region irrespective of the other leaflet's composition. (In contrast to the ``induction'' of domains described above, this would be a case of the pure leaflet's composition suppressing phase separation and thus any domain formation in the bilayer as a whole.) The size of the one-phase corners of the phase diagram in which no phase separation is possible depends quantitatively on the parameters, and they are small here due to the strong segregation.

We believe the phase diagram topology studied here (cf.\ also \cite{Williamson2014}) is qualitatively appropriate for experimental phospholipid systems:\ R-R-AR coexistence has been observed \cite{Collins2008}, implying that direct coupling should not be strong enough to eliminate it;\ AR-AR coexistence observed in molecular simulation \cite{Perlmutter2011} implies that direct coupling should not be strong enough to eliminate AR minima;\ we also note that $B < J$ is required to yield predominant pairwise antiregistration for a laterally homogeneous bilayer \cite{Williamson2014} (as measured in \cite{Zhang2004}), which places an upper limit on a realistic value of $B$ that is independent of existing estimates of the direct coupling $\gamma$ (to which $B$ can be mapped \cite{Williamson2014}). Nevertheless, it is true that both kinetics and phase diagram topology depend on the values of parameters which, at present, can only be crudely estimated in relation to real lipid systems \cite{Wagner2007, Putzel2008, Williamson2014}. Refs.~\cite{Wagner2007} and \cite{Putzel2008} provide a helpful overview of the broad possibilities of different phase diagram topologies, albeit without considering metastable states and with different (purely phenomenological) free energies to that used here. It may be useful to use molecular simulation or experiment to qualitatively deduce the appropriate phase diagram topology by exploring different overall compositions and seeing which coexistences are exhibited \cite{Collins2008}, perhaps even thereby inferring allowed ranges of model parameters. This could be complementary to the more bottom-up approach of directly estimating the microscopic model parameters \cite{Williamson2014}.

\section{Registered and antiregistered structure factors} \label{app:regantireg}

The structure within the top or bottom leaflet alone would be measured by
\begin{equation}
S_\textrm{t(b)}        (\bm{q},t)  \equiv \frac{1}{N^2} \left( \left\langle \left| \sum_{\bm{r}} \eta_\textrm{t(b)} (\bm{r},t) \exp ( i \bm{q} \cdot \bm{r}) \right|^2 \right\rangle   \right)~,
\label{eqn:Sqsingle}
\end{equation}
\noindent where angled brackets indicate averaging over independent trajectories, and
\begin{equation}
\eta_\textrm{t(b)}(\bm{r},t) =
\begin{cases}
1 & \textrm{if } S \textrm{ at } (\bm{r},t) \textrm{ in top (bottom) leaflet} \\
-1 & \textrm{if } U ~. \\
\end{cases}~
\end{equation}
\noindent Averaging over wavevectors of equal magnitude gives $S_\textrm{t(b)}({q},t)$. Such ``single-leaflet'' structure factors could in principle be experimentally obtained by deuterating or fluorescently tagging the $S$ lipid of only one leaflet. Combining the separate top and bottom structure factors together yields 
\begin{equation}
S_\textrm{single}({q},t) = \tfrac{1}{2}(S_\textrm{t}({q},t) + S_\textrm{b}({q},t))~.
\end{equation} 

$S_\textrm{single}({q},t)$ is not the same as the ``overall'' structure factor of the bilayer $S_\textrm{exp.}({q},t)$ that would be measured experimentally if a given lipid species is tagged the same in both leaflets
\begin{equation}
S_\textrm{exp.}(\bm{q},t)  \equiv \frac{1}{N^2} \left( \left\langle \left| \sum_{\bm{r}} \eta_\textrm{exp.} (\bm{r},t) \exp ( i \bm{q} \cdot \bm{r}) \right| ^2 \right\rangle   \right)~,
\label{eqn:Sq}
\end{equation}
\noindent where
\begin{equation}
\eta_\textrm{exp.}(\bm{r},t) =
\begin{cases}
2 & \textrm{if } SS \textrm{ at } (\bm{r},t) \\
1 & \textrm{if } SU \textrm{ or } US \\
0 & \textrm{otherwise}~. \\
\end{cases}
\end{equation}

\begin{figure}[floatfix]
\includegraphics[width=8.0cm]{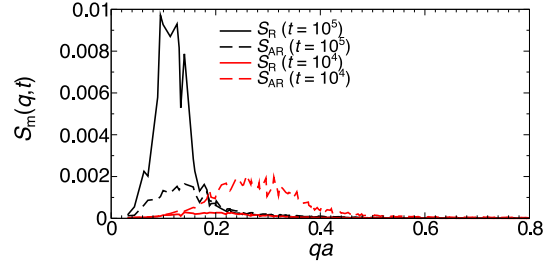}
\caption{\label{strucfac}(Color online) $J=4\,a^{-2}k_\textrm{B}T$, $B=0.48\,a^{-2}k_\textrm{B}T$, $(0.5,0.5)$ overall composition.
R and AR structure factors at the times indicated (cf.\ Fig.~\ref{0505J4B048}).}
\end{figure}

\begin{figure}[floatfix]
\includegraphics[width=8.0cm]{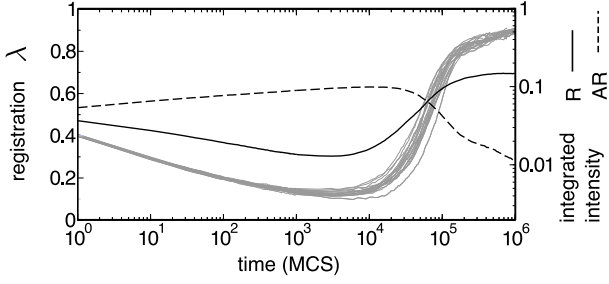}
\caption{\label{intensity}$J=4\,a^{-2}k_\textrm{B}T$, $B=0.48\,a^{-2}k_\textrm{B}T$, $(0.5,0.5)$ overall composition.
As Fig.~\ref{0505J4B048} but with black lines now showing the integrated intensity (zero'th moment) of $S_\textrm{m}({q},t)$ instead of the characteristic domain lengthscale.}
\end{figure}

We can also define separate registered and antiregistered structure factors $S_\textrm{m}(\bm{q},t)$ (where $\textrm{m} = \textrm{AR, R}$)
\begin{equation}
S_\textrm{m}(\bm{q},t)  \equiv \frac{1}{N^2} \left( \left\langle \left| \sum_{\bm{r}} \eta_\textrm{m} (\bm{r},t) \exp ( i \bm{q} \cdot \bm{r}) \right| ^2 \right\rangle   \right)~,
\label{eqn:SqRAR}
\end{equation}
\noindent where
\begin{equation}
\eta_\textrm{R}(\bm{r},t) =
\begin{cases}
1 & \textrm{if } SS \textrm{ at } (\bm{r},t) \\
-1 & \textrm{if } UU  \\
0 & \textrm{otherwise}~, \\
\end{cases}
\end{equation}
\noindent and
\begin{equation}
\eta_\textrm{AR}(\bm{r},t) =
\begin{cases}
1 & \textrm{if } SU \textrm{ at } (\bm{r},t) \\
-1 & \textrm{if } US \\
0 & \textrm{otherwise}~. \\
\end{cases}
\end{equation}
Writing $\eta_\textrm{R} = \tfrac{1}{2}(\eta_\textrm{t} + \eta_\textrm{b})$ and $\eta_\textrm{AR} = \tfrac{1}{2}(\eta_\textrm{t} - \eta_\textrm{b})$ yields 
\begin{align}
S_\textrm{R} + S_\textrm{AR} = S_\textrm{single}~,
\end{align}
\noindent i.e., the single leaflet structure factor splits into R and AR contributions. Writing $\eta_\textrm{exp.} =  \eta_\textrm{R}+1$ it can be shown that 
\begin{align}
S_\textrm{R}(\bm{q},t) \equiv S_\textrm{exp.}(\bm{q},t) \quad \forall q \neq 0~.
\end{align}
\noindent Hence, the registered and antiregistered structure factors are in principle accessible via strategic differential tagging of the two leaflets. 

From Eqs.~14 and 15, the characteristic lengthscale of R and AR domains can be monitored through time. One can also view the structure factors directly (Fig.~\ref{strucfac}) to compare the position and magnitude of the R and AR peaks, or analyse them in other ways. For example, one can consider the integrated intensity of $S_\textrm{m}({q},t)$ (Fig.~\ref{intensity}), which is the denominator of Eq.~14. This shows an initially dominant structural signal from AR domains, followed by a downturn which is accompanied by a sudden increase in the integrated intensity of $S_\textrm{R}({q},t)$ as conversion to equilibrium R-R coexistence takes place.

\begin{figure}[floatfix]
\includegraphics[width=8.0cm]{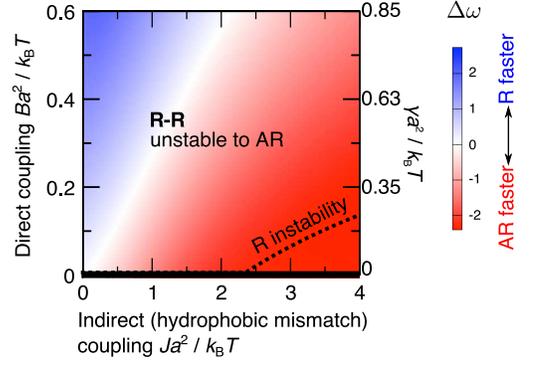}
\caption{\label{stabdiagv09}(Color online) Linear stability analysis as Fig.~\ref{stabdiag} but using $V = 0.9\,k_\textrm{B}T$ in the mean-field free energy instead of $V = 0.6\,k_\textrm{B}T$.}
\end{figure}

\section{Linear stability}\label{app:linstab}

As shown in \cite{Williamson2014}, linear stability analysis of $F_\textrm{G-L}$ (Eq.~17) yields $\omega^\textrm{R}_\textrm{max}$ and $\omega^\textrm{AR}_\textrm{max}$, the maximised (over wavenumber $q$) growth rates of the R and AR modes. Fig.~\ref{stabdiag} shows $\Delta \omega \equiv \omega^\textrm{R}_\textrm{max} - \omega^\textrm{AR}_\textrm{max}$. 
Fig.~\ref{stabdiagv09} is as Fig.~\ref{stabdiag} but using $V = 0.9\,k_\textrm{B}T$. 
The stability lines change and absolute values of the growth rates are altered, but the qualitative landscape of $\Delta \omega$ (governing which mode is fastest) is not strongly affected. 

\begin{figure}[floatfix]
\includegraphics[width=8.0cm]{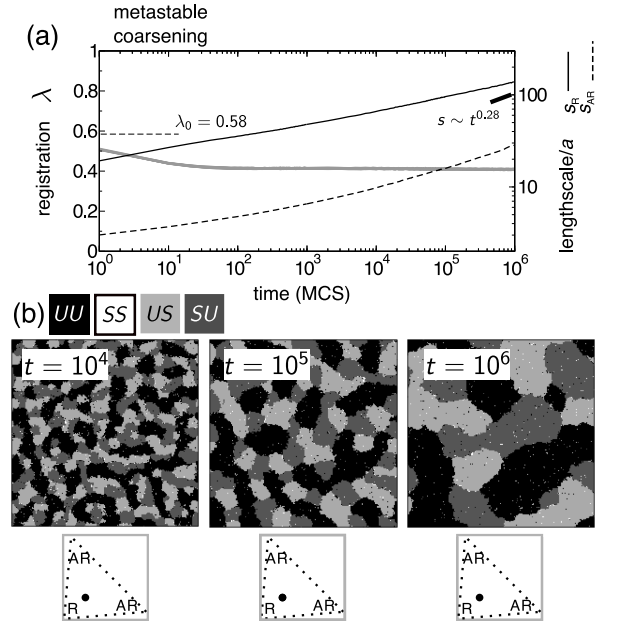}
\caption{\label{0303J4B024}$J=4\,a^{-2}k_\textrm{B}T$, $B=0.24\,a^{-2}k_\textrm{B}T$, $(0.3,0.3)$ overall composition.
As Fig.~\ref{0303J4B048}, with weaker direct coupling $B$. The bilayer is trapped in AR-AR-R coexistence.}
\end{figure}

\begin{figure}[floatfix]
\includegraphics[width=8.0cm]{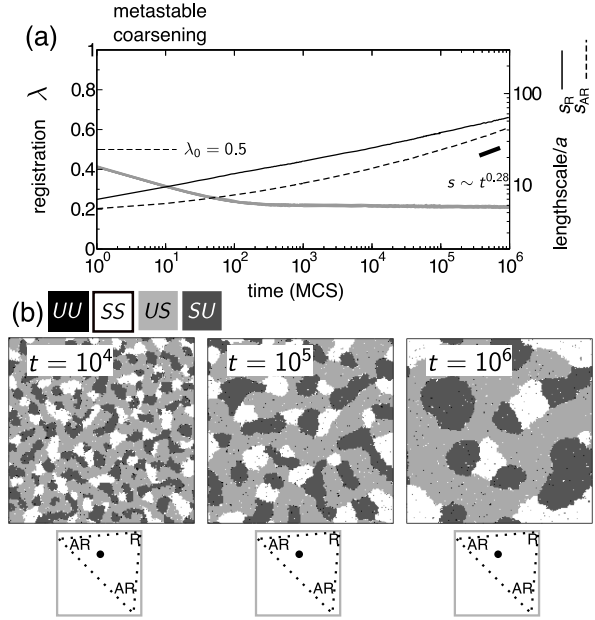}
\caption{\label{0507J4B024}$J=4\,a^{-2}k_\textrm{B}T$, $B=0.24\,a^{-2}k_\textrm{B}T$, $(0.5,0.7)$ overall composition.
As Fig.~\ref{0507J4B048}, with weaker direct coupling $B$. The bilayer is trapped in AR-AR-R coexistence.}
\end{figure}

The linear stability analysis involves a dimensionless parameter $\xi$, the relative mobility for thickness relaxation versus lateral diffusion \cite{Williamson2014}.
Figs.~4 and \ref{stabdiagv09} are calculated using $\xi = 10$ in order to approximate the value $n_r = 10$ in simulation. However, in this expected physical regime, such that thickness relaxation is much faster than lateral diffusion, the results of the stability analysis are only weakly dependent on large changes in $\xi$ \cite{Williamson2014}. 

\section{Trajectories for $(0.3,0.3)$ and $(0.5,0.7)$ with weaker direct inter-leaflet coupling}\label{app:traj}

Figs.~\ref{0303J4B024} and \ref{0507J4B024} show the effect of weaker direct coupling at $(0.3,0.3)$ and $(0.5,0.7)$ overall compositions respectively. 
Fig.~\ref{0303J4B024} is as Fig.~\ref{0303J4B048} but with smaller direct coupling $B$. This gives metastably trapped AR-AR-R coexistence instead of R-R. Fig.~\ref{0507J4B024} is as Fig.~\ref{0507J4B048} but with smaller $B$, and is trapped in metastable AR-AR-R instead of R-R-AR. Hence, metastable trapping of AR phases can occur even for overall compositions that prohibit \textit{perfect} (AR-AR) antiregistration.

\begin{acknowledgments} 
We acknowledge discussions with Lorenzo Di Michele and other members of the EPSRC CAPITALS programme grant, and thank Alexander Stukowski for the OVITO package \cite{OVITO} which was used to render simulation images.
This work was initiated at the University of Leeds, Leeds, United Kingdom, and was funded by EPSRC grant EP/J017566/1 and by Georgetown University. PDO gratefully acknowledges the support of the Ives endowment.
\end{acknowledgments} 

\bibliography{bibliography}
\end{document}